\documentclass[prb,11pt,unsortedaddress]{revtex4-1} 

\usepackage{amsmath}  
\usepackage{amsfonts} 
\usepackage{graphicx} 

\begin{document}


\title{Weak values are quantum: you can bet on it}

\author{Alessandro Romito}
\email{alessandro.romito@fu-berlin.de} 
\affiliation{Dahlem Center for Complex Quantum Systems and Fachbereich Physik, Freie Universit\"at Berlin, DE 14195 Berlin, Germany}

\author{Andrew N. Jordan}
\affiliation{Department of Physics and Astronomy \& Rochester Theory Center, University of Rochester, Rochester, New York 14627, USA}
\affiliation{Institute for Quantum Studies, Chapman University, 1 University Drive, Orange, CA 92866, USA}

\author{Yakir Aharonov}
\affiliation{School of Physics and Astronomy, Tel Aviv University, Tel Aviv, Israel}
\affiliation{Institute for Quantum Studies, Chapman University, 1 University Drive, Orange, CA 92866, USA}

\author{Yuval Gefen}
\affiliation{Department of Condensed Matter Physics, Weizmann Institute of Science, Rehovot 76100, Israel}

\date{\today}

\begin{abstract}
The outcome of a weak quantum measurement conditioned to a subsequent postselection (a “weak value” protocol) can assume peculiar values. These results cannot be explained in terms of conditional probabilistic outcomes of projective measurements. However, a classical model has been recently put forward that can reproduce peculiar expectation values, reminiscent of weak values. This led the authors of that work to claim that weak values have an entirely classical explanation. Here we discuss what is quantum about weak values with the help of a simple model based on basic quantum mechanics. We first demonstrate how a classical theory can indeed give rise to non-trivial conditional values, and explain what features of weak values are genuinely quantum. We finally use our model to outline some main issues under current research.
\end{abstract}

\maketitle 

More than 25 years ago Aharonov, Albert, and Vaidman have introduced a
measurement protocol whose outcome was termed Weak Value (WV).~\cite{Aharonov1988} This
protocol utilizes a weak measurement that avoids the collapse of the
system's wave function, conditioned on a specific outcome of a
subsequent strong (projective) measurement --termed postselection. For
a generic quantum mechanical system prepared in a state $\vert i
\rangle$, if $A$ is the operator measured weakly, and $\vert f
\rangle$--- the system’s state corresponding to the successful
postselection, the weak value obtained by the above procedure is 
$ A_w = \langle f \vert A \vert i \rangle / \langle f \vert i \rangle$.

A major feature of such a conditional measurement protocol, which has
attracted much attention, is that the weak value can be, by far,
larger than the standard quantum expectation value of the measured
observable. In a recent publication Ferrie and Combes have pointed out
that such large values may also be obtained from a strictly classical
protocol that involves conditional probabilities.~\cite{Ferrie2014} As a result, they
went on to argue that “there is nothing quantum in WVs”. Theirs is
just the latest of a series of works that, over the years, have come
up with classical analogues of the weak value protocol and associated
paradoxes (see e.g., Ref.~\onlinecite{Kirkpatrick2003}), followed by appropriate clarifications
(see e.g. Refs.~\onlinecite{Dressel2012,Ravon2007}). It appears that the question still lingers: is
it indeed true that the main features of WVs can be reproduced
following a purely classical protocol?

Well, not quite. The essential difference between classical and
quantum protocols is inherent to the fact that classically one may
acquire information on a system without disturbing it, while quantum
mechanically, measurement necessarily leads to disturbance of the
system. The implications on quantum WVs, as opposed to their classical
counterparts, are quite striking. To have a strong influence of the
post selection measurement on the initial one within a classical
protocol, the latter needs to "know" about the former. Strong
perturbation of the first measurement on the system is needed. By
contrast, large quantum WVs can be obtained for arbitrarily small
disturbance of the system by the measurement apparatus. Another
fundamental difference is that no probabilistic classical model can
consistently reproduce all post-selected outcomes of a quantum system.

\begin{figure}[h!]
\centering
\includegraphics[width=5in]{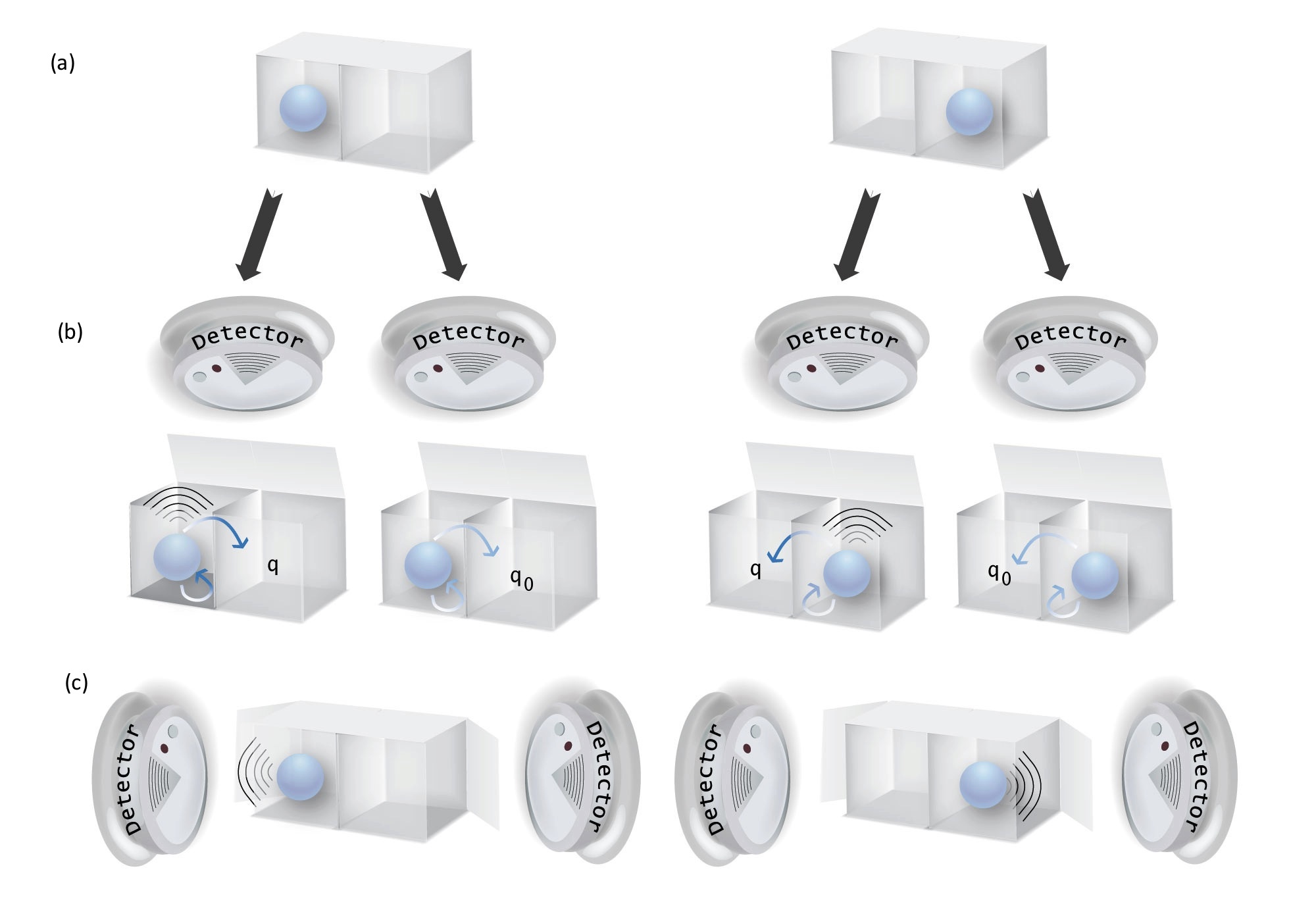}
\caption{A classical protocol for estimation of the occupation
  probabilities of box 1 and 2. (a) Preparation. Possible initial
  states of the system ---the system is prepared (with different probabilities) with either box 1 or box 2 occupied. (b) Measurement and disturbance. During the measurement a signal may be emitted (and then detected), with different probabilities, depending on which box the particle is in. The measurement is accompanied by a subsequent disturbance on the system; the particle changes its box location with probability $q$ or $q_0$, depending on whether a signal was emitted or not. (c) Post-selection. Following the measurement, the two possible final configurations (left or right) are unambiguously (strongly) detected.}
\label{fig1}
\end{figure}

To demonstrate the point of view of Ferrie and Combes, we rephrase
their simple and appealing example in the form of a parameter’s
estimation problem.~\cite{Note1} Consider a particle prepared in either of two
boxes according to a probability distribution: $p_1$ is the
probability to find it in box 1, and $p_2=1-p_1$ --- the probability
to be in box 2. The two boxes define our system. Having at our
disposal a statistical ensemble of such systems, we are interested in
determining (measuring) $p_1$ and $p_2$. This is achieved by the
following protocol (cf. Fig.~\ref{fig1}): in each repetition of the experiment
the boxes are opened and, if the particle is in box 1 it will emit a
signal with probability $P(S \vert 1) =1/2 +\lambda$;
if instead the particle is in box 2, it emits a signal with
probability $P(S \vert 2) =1/2 -\lambda$. 
The probabilities of not-emitting a signal are $P(\bar{S} \vert 1)
=1/2 -\lambda$ and $P(\bar{S} \vert 2) =1/2 +\lambda$ respectively. Note that $\lambda$ is a parameter of the detection scheme, which is
known {\it a-priori}. In fact, it controls how “good” our measurement
is: for $\lambda \to 0 $ we do not learn anything about the system and
the measurement is weak; by contrast, $\lambda=1/2$ maximizes the
information that can be extracted from the system --- the measurement
is strong. Since $p_1+p_2=1$, it is sufficient to estimate $p_1-p_2$. To do so we average the statistical data with properly assigned values of the outcome $S$,
( $\alpha_S = 1/ \lambda$) and $\bar{S}$, ($\alpha_{\bar{S}} =-1/\lambda$). The signal average is then
\begin{equation}
\langle S \rangle =\alpha_S P_S +\alpha_{\bar{S}} P_{\bar{S}},
\label{eq1}
\end{equation}
where $P_S$ and $P_{\bar{S}}$ are the probabilities of detecting or not
detecting a signal respectively, estimated simply as the number of
detected events (or non-detected events), divided by the total number
of trials. The values $\alpha_S$ and $\alpha_{\bar{S}}$ are, in fact, a
special case of so-called contextual values;~\cite{Dressel2010} they guarantee that
$\langle S \rangle = p_1-p_2 \in [-1,1]$, regardless of what the
probabilities $p_1$, $p_2$ are.
We now introduce disturbance and postselection by adding another step
to this protocol: should a signal be emitted (upon opening the boxes),
the particle has probability $q$ to switch its position between box 1
and box 2; should there be no signal emitted, switching of the
particle between the two boxes will happen with probability $q_0$. We
finally check the location of the particle. We define the conditional
signal average: this is the average $\langle S \rangle$ , conditioned
on the subsequent event f of finding the particle in box 2. It is
equal to
\begin{equation}
{}_f\langle S \rangle = \alpha_S P_{S|f} +\alpha_{\bar{S}} P_{\bar{S}
  |f} = \frac{1}{\lambda} \left( P_{S|f} -P_{\bar{S} |f } \right).
\label{eq2}
\end{equation}
Operationally, the conditional probabilities are equal to the number
of events of $S$ or $\bar{S}$, given the subsequent event $f$ (finding
the particle in box 2), divided by all trials where $f$ is found. We
define the classical analog of the weak value as
\begin{equation}
A_w^{(\textrm{classical})}=\lim_{\lambda \to \infty} {}_f\langle S \rangle.
\label{eq3}
\end{equation}
If no disturbance is introduced, this conditioned average is bounded,
$A_w^{(\textrm{classical})} \in [-1,1]$, with the same boundaries of
the unconditioned signal.

This classical protocol should be compared to its quantum analog. In
the latter, the state of the system is \emph{not} described by a
probability distribution, but rather by a density matrix. We assume
for simplicity that the system is in a pure state, $\rho=\vert i
\rangle \langle i \vert$,
with $\vert i \rangle = \sqrt{p_1} \vert 1 \rangle + \sqrt{1-p_1}
\vert 2 \rangle $. Our quantum detector, which is weakly coupled to
the system (the two boxes) with strength $\lambda$, then measures the
operator  $A= \vert 1 \rangle \langle 1 \vert - \vert 2 \rangle
\langle 2 \vert $. 
It extracts an ambiguous signal with two possible outcomes ($S$,
$\bar{S}$), analogously to the classical case (see
Ref.~\onlinecite{Ferrie2014}). Consistently assigning contextual values
($\alpha_S$,$\alpha_{\bar{S}}$) to these outcomes, the average over the
statistical data yields $\langle S \rangle = \alpha_S P_S
+\alpha_{\bar{S}} P_{\bar{S}} = \langle i \vert A \vert i
\rangle$. This is the same as the classical outcome. Now, following
this measurement we consider a unitary evolution (let it be not the
most general one), which transforms the states $\vert 1 \rangle$  and
$\vert 2 \rangle$  to  $\vert f \rangle$ and $\vert \bar{f} \rangle$,
e.g., $\vert 2 \rangle \to \vert \bar{f} \rangle = \cos(\theta/2)
\vert 1 \rangle +\sin (\theta/2) \vert 2 \rangle$, with $\theta \in
[0,2\pi]$. The conditional average of the measurement’s outcome with a
successful post-selection on box 2 is then given, in the limit of weak
measurement, $\lambda \to 0 $, by the weak value
\begin{equation}
A_w^{(\textrm{quantum})}=\lim_{\lambda \to 0} {}_f \langle S \rangle =
\frac{\langle f \vert A \vert i \rangle}{\langle f \vert i \rangle} =
 \frac{\sqrt{p_1 \sin(\frac{\theta}{2})} +\sqrt{
     (1-p_1)\cos(\frac{\theta}{2}) } }{\sqrt{p_1
     \sin(\frac{\theta}{2})} -\sqrt{  (1-p_1)\cos(\frac{\theta}{2}) } }.
\label{eq4}
\end{equation}
The weak value obtained for any pre- and post-selection in the quantum
model can be reproduced by the conditional outcome of the classical
model, i.e. $A_w^{(\textrm{quantum})}= A_w^{(\textrm{classical})}$, upon choosing specific initial state and parameters. This has led Ferrie and Combes to conclude that ``weak values are not quantum''. The simple example
used to put forward their claim~\cite{Ferrie2014} employed $\sqrt{p_1}
=\cos(\theta/2)$, hence $A_w^{(\textrm{quantum})} = 1/\cos(\theta)$. 
The same large number is reproduced through the classical model in the
weak coupling limit  ( $\lambda \to 0 $ ) by choosing $p_1=1$,
$q_0=(\cos\theta -\lambda)/(1-\lambda)$, $q=(\cos \theta +\lambda)/
(1+\lambda)$.

This example, following the footsteps of several others~\cite{Kirkpatrick2003,Dressel2012}, unambiguously demonstrates that the mere fact that weak values can be arbitrarily large is not per se a unique feature of quantum mechanics. In what way is then the weak value a genuine quantum mechanical entity? First of all, no classical model can consistently reproduce the weak values of a quantum system and all post-selected states. In fact,
as nicely highlighted in Ref.~\onlinecite{Dressel2015,Pusey2014}, were
this the case in a classical protocol, along with large weak values,
some post-selection events would have to occur with negative
probabilities!!

A more quantitative way for discerning the quantum nature of weak
values is to focus on the detector’s back-action on the system. In the
quantum case, the weaker the measurement  ($\lambda \to 0$), the
lesser is the postselection probability modified from its unperturbed
value; nonetheless, large weak values can emerge.  This is to be
contrasted with a classical protocol, also in the weak measurement
limit. In order to obtain the same large value (now of
$A_w^{(\textrm{classical})}$), one needs to modify the post-selection
probability substantially.~\cite{Dressel2015,Hoffmann2014} In the simple classical model outlined
by Ferrie and Combes, the postselection probability in the absence of
measurement is identically zero; when performing the weak measurement
the change in the post selection probability is $P(f) =P(f \vert S)
+P(f \vert \bar{S}) \sim \cos \theta \gg \lambda$. The qualitative
difference with the infinitesimal change of post-selection probability
in the quantum case is evident, and may be put on a quantitative
basis.~\cite{Ipsen2015} One can look at the minimal back-action for
fixed information acquired by the detector in the classical and
quantum cases. Trivially, in the former it is possible to attain zero
backaction; in the quantum case a finite minimal back-action is
unavoidable. An important observation is that one obtains the
Aharonov-Albert-Vaidman weak value as the weak measurement limit of a
conditioned average in a wide class of models, where the detector
minimally disturbs the system.~\cite{Dressel2012,Ipsen2015} How to
define weak values as minimal backaction limit over a wide family of
quantum measurement protocols is the subject of current research.

It is clear from the above that, while anomalous classical conditional values imply a large disturbance, and they depend on the (arbitrarily chosen) value of the latter, quantum weak values conform to the minimum back-action requirement. In fact,
postselected measurements with minimum back-action offer an operational approach to deal with conditional averages in the quantum realm. The related weak values have been used as a new tool to define physical quantities not amenable to simple, direct strong measurement (e.g., tunneling time,~\cite{Steinberg1995} many-body cotunneling time~\cite{Romito2014}).
In short, not only are weak values quantum, but they also constitute an interesting tool to explore hidden facets of quantum mechanics. Much of this remains a challenge for the future.



\end{document}